\documentclass[12pt,twoside]{article}
{\catcode `\@=11 \global\let\AddToReset=\@addtoreset}
\AddToReset{equation}{section}
\renewcommand{\theequation}{\thesection.\arabic{equation}}  
\def\greaterthansquiggle{\raise.3ex\hbox{$>$\kern-.75em\lower1ex\hbox{$\sim$}}}
\def\lessthansquiggle{\raise.3ex\hbox{$<$\kern-.75em\lower1ex\hbox{$\sim$}}}
\newcommand{\beq}{\begin{equation}}
\newcommand{\eeq}{\end{equation}}
\newcommand{\beqa}{\begin{eqnarray}}
\newcommand{\eeqa}{\end{eqnarray}}
\newcommand{\beqan}{\begin{eqnarray*}}
\newcommand{\eeqan}{\end{eqnarray*}}
\newcommand{\ba}{\begin{array}}
\newcommand{\ea}{\end{array}}
\newcommand{\no}{\nonumber}

\newcommand{\sgn}{{\rm sgn}}

\newcommand{\Un}{\underline}
\newcommand{\ol}{\overline}
\newcommand{\ra}{\rightarrow}

\newcommand{\ve}{\varepsilon}
\newcommand{\vp}{\varphi}

\newcommand{\wt}{\widetilde}

\newcommand{\A}{{\cal A}}
\newcommand{\B}{{\cal B}}
\newcommand{\C}{{\cal C}}
\newcommand{\D}{{\cal D}}

\newcommand{\F}{{\cal F}}

\newcommand{\Ha}{{\cal H}}

\newcommand{\T}{{\cal T}}

\newcommand{\V}{{\cal V}}

\newcommand{\st}{\stackrel}
\newcommand{\dfrac}{\displaystyle \frac}
\newcommand{\dint}{\displaystyle \int}
\newcommand{\dsum}{\displaystyle \sum}
\newcommand{\dprod}{\displaystyle \prod}

\newcommand{\dlim}{\displaystyle \lim}

\def\nz{\ifmmode {I\hskip -3pt N} \else {\hbox {$I\hskip -3pt N$}}\fi}
\def\zz{\ifmmode {Z\hskip -4.8pt Z} \else
       {\hbox {$Z\hskip -4.8pt Z$}}\fi}
\def\qz{\ifmmode {Q\hskip -5.0pt\vrule height6.0pt depth 0pt
       \hskip 6pt} \else {\hbox
       {$Q\hskip -5.0pt\vrule height6.0pt depth 0pt\hskip 6pt$}}\fi}
\def\rz{\ifmmode {I\hskip -3pt R} \else {\hbox {$I\hskip -3pt R$}}\fi}
\def\cz{\ifmmode {C\hskip -4.8pt\vrule height5.8pt\hskip 6.3pt} \else
       {\hbox {$C\hskip -4.8pt\vrule height5.8pt\hskip 6.3pt$}}\fi}

\def\lint{\int\limits}
\normalsize
\begin{document}
\bibliographystyle{plain}
\begin{titlepage}
\begin{flushright}
UWThPh-1999-37\\
ESI-720-1999\\
math-ph/9906020\\ 
June 1, 1999  
\end{flushright}
\vspace*{0.3cm} 
\begin{center}
{\Large \bf  
Anyons and the Bose-Fermi duality\\[10pt] 
in the finite-temperature Thirring model$^\star$}\\[25pt]
N. Ilieva$^{\ast,\sharp}$ and W. Thirring  \\ [8pt]
Institut f\"ur Theoretische Physik \\ Universit\"at Wien\\
\smallskip
and \\
\smallskip
Erwin Schr\"odinger International Institute\\
for Mathematical Physics\\
\end{center}

\vspace{0.4cm}
\begin{abstract}
Solutions to the Thirring model are constructed in the framework of
algebraic quantum field theory. It is shown that for all positive
temperatures there are fermionic solutions only if the coupling constant
is $\lambda = \sqrt{2(2n+1)\pi}, \, n\in \bf N$. These fermions are
inequivalent and only for $n=1$ they are canonical fields. In the general
case solutions are anyons. Different anyons (which are uncountably many)
live in orthogonal spaces and obey dynamical equations (of the type of
Heisenberg's ``Urgleichung") characterized by the corresponding values of
the statistic parameter. Thus statistic parameter turns out to be related
to the coupling constant $\lambda$ and the whole Hilbert space becomes
non-separable with a different ``Urgleichung" satisfied in each of its
sectors. This feature certainly cannot be seen by any power expansion in
$\lambda$. Moreover, since the latter is tied to the statistic parameter,
it is clear that such an expansion is doomed to failure and will never
reveal the true structure of the theory.

The correlation functions in the temperature state for the canonical
dressed fermions are shown by us to coincide with the ones for the bare
fields, that is in agreement with the uniqueness of the $\tau$-KMS state
over the CAR algebra ($\tau$ being the shift automorphism). Also the
$\alpha$-anyon two-point function is evaluated and for scalar field it
reproduces the result that is known from the literature.

\vspace{6pt}
PACS codes: 03.70.+k, 11.10.Wx, 11.10.Kk, 05.30.-d

\end{abstract}

\vspace{8pt}
\vfill
{\footnotesize

$^\star$ Work supported in part by ``Fonds zur F\"orderung der
wissenschaftlichen Forschung in \"Osterreich" under grant P11287--PHY, 
{\sl to be published in} {\it Theor. Math. Phys.}

$^\ast$ On leave from Institute for Nuclear Research and Nuclear Energy,
Bulgarian Academy of Sciences, Boul.Tzarigradsko Chaussee 72, 1784 Sofia,
Bulgaria

$^\sharp$ E--mail address: ilieva@ap.univie.ac.at}

\end{titlepage}

\begin{quotation}
{\it It is an honour for us to present this paper to the 30$^{{\small
\,th}}$ anniversary of ``Theoretical and Mathematical Physics". With its
scientific reputation and high standards the journal takes a valuable
place in the heritage of its founder, N.N.~Bogoliubov --- one of the
outstanding scientists of this century, and we wish this to be kept also
in the future.} 
\end{quotation}

\section{Introduction}
After T.D. Lee had constructed a model of a soluble QFT \cite{TDL} many people
tried to find other examples; but to solve a nontrivial relativistic QFT seemed
out of the question. The idea that Bethe's ansatz \cite{B} could be
successfully used to solve also Heisenberg's ``Urgleichung" \cite{Hei} reduced 
to one
space one time dimension then led to a soluble relativistic field theory
--- the Thirring model \cite{WT}. During the years, this model has not
only been 
extensively studied but has also been actively used for analysis, testing
and illustration of various phenomena in two-dimensional field theories. 

It is not our purpose to review the enormous literature on the subject but we
rather focus on the very starting point --- Heisenberg's Urgleichung.
With no bosons present in it at all, it represents
the ultimate version of the opinion  that fermions should enter the basic 
formalism of the fundamental theory of elementary particles that is usually 
taken for granted.
 
The opposite point of view, namely that a theory including only
observable fields, necessarily uncharged bosons, is capable of describing
evolution and symmetries of a physical system, being the kernel of algebraic
approach to QFT \cite{H}, also enjoys an enthusiastic support. 
Actually, the question which is thus posed and which is of principal
importance is whether and in which cases definite conclusions about the
time evolution and symmetries of charged fields can be drawn from the
knowledge about the observables that is gained through experiment.

As we will see, there is no possibility to judge this matter on the basis of
the model in question, since both  formulations can be equally well used to
construct the physically relevant objects --- the dressed fermions. 

In any case, before claiming that an ``Urgleichung" of the type
\beq
\not\!\partial\psi(x) = \lambda\psi(x)\bar\psi(x)\psi(x)
\eeq
determines the whole Universe one should see whether it determines anything
mathematically and it is our aim in the present paper to discuss the
elements
needed to make its solution well defined. In fact we shall first consider only
one chiral component and we shall restrict ourselves to the 
two-dimensional spacetime, so that this component depends only on one light
cone coordinate. Also the bose-fermi duality takes place there and we
want to make use of it. This phenomenon amounts to the fact that in
certain models formal functions of fermi fields can be written that have
vacuum expectation values and statistics of bosons and vice versa.
The equivalence is understood within perturbation theory: the perturbation
series for the so-related theories are term-by-term equivalent (they
may perfectly well exist even if the models are not exactly solvable or if
their physical sensibility is doubtful).

There are two facts which make such a duality possible. First comes the
main reason why soluble fermion models exist in two dimensions, that is
that fermion currents can be constructed as ``fields" acting on the
representation space for the fermions. Also, the ``bosons into fermions"
programme rests on the fact that bosons in question are just the currents
and fermions are essentially determined by their commutation relations
with them. Second comes the observation which has been made in the
pioneering works by Jordan \cite{J} and Born \cite{BNN}: due to the
unboundedness from below of the free-fermion Hamiltonian the fermion
creation and annihilation operators must undergo what we should call now a
Bogoliubov transformation. Thus the stability of the system is achieved
but in addition an anomalous term (later called ``Schwinger term")
appeares in the current commutator, that in turn enables the 
``bosonization".

The bose-fermi duality is actually well established when the construction of
bosons out of fermions is considered so that consistent expressions exist
for the fermion bilinears that are directly related to the observables of
the theory.

The problem of rigorous definitions of operator valued distributions and
eventually operators having the basic properties of fermions by taking
functions of bosonic fields is rather more delicate. On the level of operator
valued distributions solutions have been given by Dell'Antonio et al.\cite{DA}
and Mandelstam \cite{M} and on the level of operators in a Hilbert space --- by
Carey and collaborators \cite{CR, CHB} and in a Krein space by Acerbi, Morchio
and Strocchi \cite{AMS}.

Thus our goal is to give in one and the same setting a precise meaning to
the following three ingredients
\beq
\begin{array}{llcl}
(a) & [\,\psi^\ast(x), \psi(x')\,]_+ = \delta(x\!-\!x'), \qquad 
[\,\psi(x), \psi(x')\,]_+ = 0 & \qquad & {\rm CAR} \\[7pt]
(b) & j(x) = \psi^\ast(x)\psi(x) & \qquad & {\rm Current}\\[7pt]
(c) & \frac{1}{i}\frac{d}{dx}\psi(x) = \lambda j(x)\psi(x) & 
\qquad & {\rm Urgleichung}\\
\end{array}
\eeq

We shall approach it by constructing a series of algebraic inclusions,
starting from the CAR-algebra of bare fermions. Eq.(1.2c) involves
(derivatives of) objects which are according to (1.2a) rather
discontinuous. Therefore it is expedient to pass right away to the level of
operators in Hilbert space since the variety of topologies there provides a
better control over the limiting procedures. In general norm convergence
can hardly be hoped for but we have to strive at least for strong
convergence such that the limit of the product is the product of the
limits. With $\psi_f = \int_{-\infty}^{\infty}dx f(x)\psi(x)$, (1.2a)
becomes
\beq
[\,\psi^\ast_f, \psi_g\,]_+ = \langle f\vert g\rangle
\eeq
for $f\in L^2(\bf R)$ and $\langle .\vert .\rangle$ the scalar product in
$L^2(\bf R)$. This shows that $\psi_f$'s are bounded and form the
$C^\ast$-algebra CAR. There the translations $\,x\ra x+t\,$ give an
automorphism
$\tau_t$ and we shall use the corresponding KMS-states $\omega_\beta$ and the
associated representation $\pi_\beta$ to extend CAR. Though there $j = \infty$,
one can give a meaning to $j$ as a strong limit in $\Ha_\beta$ by smearing
$\psi(x)$ over a region $\ve$ to $\psi_\ve(x)$ and then defining
$$
j_f = \int dx f(x) \lim_{\ve\to 0}\left(\psi^*_\ve(x)\psi_\ve(x) -
\omega_\beta(\psi^*_\ve(x)\psi_\ve(x))\right), \qquad f:{\bf R}\to {\bf R}
$$
These limits exist in the strong resolvent sense and define self-adjoint
operators with a multiplication law
\beq
e^{ij_f}e^{ij_g} = e^{\frac{i}{8\pi}\int
dx(f(x)g'(x)-f'(x)g(x))}e^{ij_{f+g}}\, .
\eeq
Thus the current algebra $\A_c$ is determined. Its Weyl structure is the
same for all positive $\beta$ and $\omega_\beta$ extends to $\A_c$.

To construct the interacting fermions which on the level of distributions look
like 
$$
\Psi(x) = Z \,e^{i\lambda\int_{-\infty}^{x}dx'j(x')} \st{?}{=}
\lim_{\ve \ra 0_+} \lim_{R \ra \infty} \Psi_{\ve, R}(x)
$$ 
(with some renormalization constant $Z$) poses both infrared ($R \ra \infty$) 
and ultraviolet ($\ve \ra 0$) problems. For
$$
\Psi_{\ve, R}(x) = e^{i\lambda\int dx'(\vp_\ve (x-x')-\vp_\ve (x-x'+R))
j(x')}, \qquad 
\vp_\ve(x) := \left\{\ba{cl} 1 & \mbox{ for } x \leq -\ve \\
-x/\ve & \mbox{ for } - \ve \leq x \leq 0 \\
0 & \mbox{ for } x \geq 0 \ea \right.
$$
neither the limit $R\to\infty$ nor the limit $\ve\to 0$ exist even as weak 
limits in $\Ha_\beta$. So an extension of $\pi_\beta(\A_c)''$ is needed to
accomodate such a kind of objects.

There are two equivalent ways of handling the infrared problem. Since the
automorphism generated by the unitaries $\Psi_{\ve, R}(x)$ for 
$R\to\infty$ converges to a limit $\gamma$, one can form with it the
crossed product  $ \bar\A_c = \A_c \,\st{\gamma}{\bowtie}\,\rm Z$, so that
in $\bar\A_c$ there are unitaries with the properties which the limit
should have \cite{IN, IT}. On the other hand, the symplectic form in
(1.4) and the state $\omega_\beta$ can be defined for the limiting
element $\Psi_\ve(x)$ and this we shall do in what follows. The former
route will be then discussed in Appendix A.

In  any case $\bar\Ha_\beta$ assumes a sectorial structure, the subspaces  
$\A_c\,\dprod_{i=1}^{n} \Psi_\ve(x_i)\vert\Omega\rangle\,$ for different
$n$ are orthogonal and thus may be called $n$-fold charged sectors. The
$\Psi_\ve(x)$'s have the property that for $\vert x_i - x_j\vert > 2\ve$
they obey anyon statistics with parameter $\lambda^2$ and an Urgleichung
(1.2c) where $j(x)$ is averaged over a region of lenght $\ve$ below $x$. 

Then, by removing the ultraviolet cut-off 
the sectors abound and the subspaces
$\,\A_c\Psi(x)\vert\Omega\rangle\,$ become orthogonal for different $x$, so
$\bar\Ha_\beta$ becomes non-separable. To get canonical fields of the type
(1.3) one has to combine $\ve\ra 0_+$ with a field renormalization
$\Psi_\ve \to \ve^{-1/2}\Psi_\ve$ such that
$$ 
\lim_{\ve\ra 0_+}\ve^{-1/2}\int dx f(x)\Psi_\ve(x) = \Psi_f 
$$
converge strongly in $\bar\Ha_\beta$ and satisfy (1.2c) in sense of
distributions.

The current (1.2b) has been constructed with the bare fermions $\psi$ and
is obviously sensitive to the infinite renormalization in the dressed
field $\Psi$. Therefore it is better to replace (1.2b) by the requirement
that $j_f$ generates a local gauge transformation. Indeed,
\beq
e^{ij_f}\Psi_g e^{-ij_f} = \Psi_{e^{if}g}
\eeq
holds and in this sense (1.2b) is also satisfied.

However, the objects so constructed are in general anyons and only for
particular values of the coupling constant, $\lambda = 
\sqrt{2(2n+1)\pi},\,\, n\in\bf N$, they are fermions, so that the coupling
constant is tied to the statistic parameter. Thus we find that there is
indeed some magic about the Urgleichung inasmuch as on the quantum level it
allows fermionic solutions by this construction only for isolated values of the coupling
constant $\lambda$ whereas classically $\Psi(x) = Z \,e^{i\lambda\int_
{-\infty}^{x}dx'j(x')}$ solves (1.2c) for any $\lambda$. This feature can 
certainly not be seen by any power expansion in $\lambda$.

The scheme presented here means, that the dressed fermions obtained
for special values of $\lambda$ (and distinct from the bare ones) can be
constructed either from bare fermions or directly from the current
algebra, so no priority might be asigned to either of the two formulations,
bosonic or fermionic. To make this statement precise, the correlation
functions arrising in both cases have to be compared. As we shall see, for
canonical fermions they do coincide that is in agreement with the
uniqueness of the $\tau$-KMS state over the CAR algebra. We shall also
discuss the thermal correlators for the anyonic fields, in particular, we
shall find an agreement with the recent result in \cite{BY} for the
scalar-field case.

By a {\it symmetry} of a physical system an automorphism $\,\alpha\,$ of 
the algebra $\,\A\,$ which describes it is understood. The algebraic
chain of inclusions we construct gives an example of a {\it symmetry
destruction}, that is, for a given extension $\,\B\,$ of the algebra
$\,\A\,$, $\,\B \supset \A$, $\,\not\hspace{-1mm}\exists \beta \in
\mbox{Aut } \B: \,\, \beta\vert_\A = \alpha$ for some $\,\alpha \in
\mbox{Aut }\A$. This phenomenon is related to the spontaneous collapse of
a symmetry \cite{BO} and  in contrast to the spontaneous symmetry breaking
\cite{NT}, it cannot occur in a finite-dimensional Hilbert space.
\vspace{0.6cm}

\section{Bosons out of fermions: the CAR-algebra, its KMS-states and associated 
v. Neumann algebras}
Let us consider the  C*-algebra $\A^l$ formed by the bounded operators
\beq
\psi_f = \int_{-\infty}^\infty dx \psi(x) f(x) =
\int_{-\infty}^\infty \frac{dp}{2\pi} \wt \psi(p) \wt f(p), \qquad
\wt f(p) = \int_{-\infty}^\infty dx \; e^{ipx} f(x)
\eeq
with $\psi(x)$, $x \in {\bf R}$, being operator-valued distributions
which satisfy
\beq
[\psi^*(x),\psi(x')]_+ = \delta(x\!-\!x'),
\eeq
so, describing the left movers (we have asigned a superscript to the
relevant
quantities, $x$ stands for $x-t$) and $f \in L^2({\bf R})$. This algebra is 
characterized by
\beq
[\psi^*_f,\psi_g]_+ = \langle f|g\rangle = \int dx f^*(x) g(x).
\eeq
Translations $\tau_t$ define an automorphism of $\,\A^l\,$
\beq
\tau_t \psi_f = \psi_{f_t}, \quad
f_t(x) = f(x-t).
\eeq
$\A^l\,$ inherits the norm from $L^2({\bf R})$ such that $\,\tau_t\,$ is
(pointwise) normcontinuous in $\,t\,$ and even normdifferentiable for the
dense set of $f$'s for which
$$
\lim_{\delta \ra 0_+} \frac{f(x+\delta)-f(x)}{\delta} = f'(x)
$$
exists in $L^2({\bf R})$
\beq
\left.\frac{d}{dt} \; \tau_t \psi_f \right|_{t=0} = - \psi_{f'}.
\eeq
The $\tau$-KMS-states over $\A^l\,$ are given by
\beq
\omega_\beta(\psi^*_f \psi_g) = 
\int_{-\infty}^\infty \frac{dp}{2\pi} \frac{\wt f^*(p) \wt g(p)}
{1 + e^{\beta p}} =
\sum_{n=-\infty}^\infty \frac{(-1)^n}{2\pi}
\int \frac{dx dx' f^*(x) g(x')}{i(x-x') - n\beta + \ve}, \qquad
\ve \ra 0_+,
\eeq
$$
\omega_\beta(\psi_g \psi^*_f) = 
\omega_\beta(\psi^*_f \tau_{i\beta} \psi_g).
$$
With each $\omega_\beta$ are associated a representation $\pi_\beta$ with
cyclic vector $|\Omega\rangle\,$, $\,\omega_\beta(a) = \langle \Omega|a|\Omega
\rangle$ in $\,\Ha_\beta = \ol{\A^l|\Omega\rangle}\,$ and a v.~Neumann algebra
$\,\pi_\beta(\A^l)''$. It contains the current algebra $\A^l_c$ which gives
the formal expression $j(x) = \psi^*(x) \psi(x)$ a precise meaning.

To show this, let us recall two lemmas (for the proofs see \cite{IT}) which make
the whole construction transparent:
\paragraph{Lemma} (2.7) \\
If the kernel $K(k,k'): {\bf R}^2 \ra C$ is as operator $\geq 0$ and
trace class $(K(k,k) \in L^1({\bf R}))$, then $\forall \; \beta \in
{\bf R}^+$
\beqan
\lim_{M \ra \pm \infty} B_M &:=& \lim_{M \ra \pm \infty}
\frac{1}{(2\pi)^2}\int dk dk' K(k,k') \wt \psi^*(k+M) \wt \psi(k'+M) = \\
&=& \frac{1}{(2\pi)^2}\int dk dk' \lim_{M \ra \pm \infty} K(k,k') \omega_\beta
(\wt \psi^*(k+M) \wt \psi(k'+M)) = \\
&=& \left\{\ba{cl} \frac{1}{2\pi}\int dk \; K(k,k) & \mbox{ for } M \ra +\infty \\
0 & \mbox{ for } M \ra - \infty \ea \right.
\eeqan
in the strong sense in $\Ha_\beta$. \\[3pt]

However, if $\int |K|^2$ keeps increasing with $M$, then 
$B_M - \langle B_M\rangle$ may nevertheless tend to an (unbounded) operator.

\paragraph{Lemma} (2.8) \\
If
$$
B_M = \frac{1}{(2\pi)^2}\int dk dk' \wt f(k-k') \Theta(M-|k|) 
\Theta(M-|k'|) \wt \psi^*(k) \psi(k')
$$
with $\wt f$ decreasing faster than an exponential and being the Fourier 
transform 
of a positive function, then the difference $\,B_M - \omega_\beta(B_M)\,$ is a 
strong Cauchy sequence for $\,M \ra \infty\,$ on a dense domain on 
$\Ha_\beta$.

\paragraph{Remarks} (2.9)
\begin{enumerate}
\item (2.7) substantiates the feeling that for $k > 0$ most levels are
empty and for $k < 0$ most are full.
\item $B_M$ is a positive operator and by diagonalizing $K$ one sees
$$
\|B_M\| = \|K\|_1 = \frac{1}{2\pi}\int dk\; K(k,k).
$$
\item As just mentioned, $\,\|B_M\| < 2M \wt f(0)\,$ and $f(x) \geq 0$ is not 
a serious restriction since any function is a linear combination of positive
functions.
\item Since the limit $j_f$ is unbounded the convergence is not on all
of $\Ha_\beta$, however since for the limit $j_f$ holds $\tau_{i\beta}j_f = 
j_{e^{\beta p} f}$,  the dense domain is invariant under $j_f$. Thus we have
strong resolvent convergence which means that bounded functions of
$B_M$ converge strongly. Also the commutator of the limits is the limit
of the commutator.
\end{enumerate}
\vspace{0.4cm}

Thus we conclude that the limit exists and is selfadjoint on a suitable domain.
We shall write it formally
$$
\ba{rcl}
j_f &=& \dlim_{M \ra \pm \infty}
\dfrac{1}{(2\pi)^2}\int_{-\infty}^\infty dk dk'\, K(k,k') \wt \psi^*(k+M)
\wt \psi(k'+M)
\\[8pt]
 &=& \dfrac{1}{(2\pi)^2}\int_{-\infty}^\infty dk dk \;
\wt f(k-k')
: \wt \psi(k)^* \wt \psi(k'): 
\ea \eqno(2.10)
$$

Next we show that the currents so defined satisfy the CCR with a suitable symplectic form
$\sigma$ \cite{J, Schwinger}.
\paragraph{Theorem} (2.11)
$$
[j_f,j_g] = i \sigma(f,g) =
\int_{-\infty}^\infty \frac{dp}{(2\pi)^2} \; p \wt f(p) \wt g(-p) =
\frac{i}{4\pi} \int_{-\infty}^\infty dx(f'(x)g(x) - f(x)g'(x)).
$$
\paragraph{Proof:} For the distributions $\wt \psi(k)$ we get algebraically
$$
[\wt\psi^*(k) \wt\psi(k'),\wt\psi^*(q)\wt\psi(q')] =
2\pi\left[\wt\psi^*(k)\wt\psi(q')\delta(q\!-\!k') -
\wt\psi^*(q)\wt\psi(k') \delta(k\!-\!q')\right]
$$
and for the operators after some change of variables
$$
\frac{1}{(2\pi)^3}\int dkdpdp' \wt f(p) \wt g(p') \wt\psi^*(p+p'+k) \wt\psi(k)
\Theta(M-|k|) \Theta(M-|p+p'+k|) \cdot
$$
$$
\cdot \left[\Theta(M-|p'+k|) - \Theta(M-|p+k|)\right].
$$
For fixed $p$ and $p'$ and $M \ra \infty$ we see that the allowed region
for $k$ is contained in $(M-|p|-|p'|,M)$ and
$(-M,-M+|p|+|p'|)$. Upon $k \ra k \pm M$ we are in the situation of (2.7),
thus we see that the commutator of the currents (2.10)
is bounded uniformly in $M$ if $\wt f$ and 
$\wt g$ decay faster than exponentials and converges to the expectation
value. This gives finally
$$
\int_{-\infty}^\infty \frac{dp}{(2\pi)^2} \; \wt f(p) \wt g(-p)
\int dk \; \Theta(M-|k|) \left[\Theta(M - |k-p|) - \Theta(M - |k+p|)\right]
\frac{1}{1 + e^{\beta k}}
$$
$$
\st{M \ra \infty}{\longrightarrow} \int_{-\infty}^\infty \frac{dp}{(2\pi)^2}
\; p \wt f(p) \wt g(-p).
$$

\paragraph{Remarks} (2.12)
\begin{enumerate}
\item Since the $j_f$'s satisfy the CCR they cannot be bounded and it is
better to write (2.11) in the Weyl form for the associated unitaries
$$
e^{ij_f} \; e^{ij_g} = e^{\frac{i}{2} \sigma(g,f)} \; e^{ij_{f+g}} =
e^{i\sigma(g,f)} \; e^{ij_g} \; e^{ij_f}.
$$ 
\item The currents $j_f$ are selfadjoint, so the unitaries $\,e^{i\alpha j_f}$ 
generate one-parameter groups --- the local gauge transformations
$$
e^{-i\alpha j_f} \; \psi_g \; e^{i\alpha j_f} = \psi_{e^{i\alpha f}g}.
$$
\item The state $\,\omega_\beta\,$ can be extended to $\,\bar \omega_\beta\,$
over $\,\pi_\beta(\A^l)''$ and $\tau_t$ to $\bar \tau_t,\,\,
\bar \tau_t \in \mbox{Aut }\pi_\beta(\A^l)''$ with $\bar \tau_t \,j_f = j_{f_t}$.
Furthermore $\,\bar \omega_\beta\,$ is $\,\bar \tau$-KMS and is calculated to 
be (\cite{IT}, see also \cite{BY})
$$
\bar \omega_\beta(e^{ij_f}) = \exp \left[ -\frac{1}{2} \int_{-\infty}^\infty
\frac{dp}{(2\pi)^2} \frac{p}{1 - e^{-\beta p}} |\wt f(p)|^2  \right].
$$
\item A physically important symmetry of the algebra $\,\A^l\,$, the parity $P$,
$$ 
P \in \mbox{Aut }\A^l, \quad  P\psi_f = \psi_{Pf}, \quad P f(x) = f(-x)
$$
is destroyed in $\pi_\beta$, since 
$$
[j(x),j(x')] = -\,\frac{i}{2\pi}\,\delta'(x\!-\!x')
$$
is not invariant under $j(x) \ra j(-x)$. 
Thus $P \notin \mbox{Aut }\pi_\beta(\A^l)''$ and $\,\bar\omega_\beta$ is not
P-invariant. 
\item The extended shift automorphism $\,\bar \tau_t\,$ is not only strongly 
continuous but for
suitable $f$'s also differentiable in $\,t\,$ (strongly on a dense set in 
$\Ha_\beta$)
$$
\frac{1}{i} \frac{d}{dt} \bar \tau_t e^{ij_f} = 
\left[ j_{f'_t} + \frac{1}{2} \sigma(f_t,f'_t)  \right] e^{ij_{f_t}} =
e^{ij_{f_t}} \left[j_{f'_t} - \frac{1}{2} \sigma(f_t,f'_t) \right] =
\frac{1}{2} \left[ j_{f'_t} \; e^{ij_{f_t}} + e^{ij_{f_t}} j_{f'_t} \right].
$$
\item The symplectic structure is formally independent on $\beta$ \cite{HG}, 
however for $\beta < 0$ it changes its sign, $\sigma \to -\sigma$, and for 
$\beta = 0$ (the tracial state) it becomes zero.
\end{enumerate}

Thus starting from a CAR-algebra $\A^l$, we identified in $\pi_\beta(\A^l)''$
bosonic fields --- the currents, which satisfy CCR's. The crucial
ingredient
needed was the appropriately chosen state. Here we have used the KMS-state
(which is unique for the CAR algebra). Another possibility would be to
introduce the Dirac vacuum (filling all negative energy levels in the Dirac
sea). This is  what has been done in the thirties \cite{J, BNN}, in order
to achieve stability for a fermion system, and recovered later by Mattis and
Lieb \cite{ML} in the context of the Luttinger model. Thus as an additional 
effect the appearance of an anomalous term in the current commutator (later 
called Schwinger term) had been discovered that actually enables bosonization 
of these two-dimensional models.

\vspace{0.6cm}

\section{Extensions of $\A_c$: fermions out of bosons}
So far $\A^l_c$ was defined for $j_f$'s with $\, f \in C_0^\infty$, for
instance. The algebraic structure is determined by the symplectic form 
$\sigma(f,g)$ (2.11) which 
is actually well defined also for the 
Sobolev space, $\sigma(f,g) \ra \sigma(\bar f, \bar g), \, \bar f, \bar g \in
H_1, \, H_1 = \{f : f,f' \in L^2\}\,$. Also $\bar \omega_\beta$
can be extended to $H_1$, since $\,\bar \omega_\beta(e^{ij_{\bar f}}) > 0\,$
for $\,\bar f \in H_1$.
The anticommuting operators we are looking for are of the form
$e^{ij_f}$, with $f(x) = 2\pi \Theta(x_0\!-\!x) \not\in H_1$. Still one can give
$\sigma(f,g)$ a meaning for such an $f$. However, the corresponding state 
$\omega_\beta$ exhibits singular behaviour for both $p \ra 0$ and 
$p \ra \infty$, so that 
$$
\omega_\beta(e^{ij_\Theta}) = \exp \left[-\frac{1}{2} \int_{-\infty}^\infty
\frac{dp}{p(1 - e^{-\beta p})} \right] = 0
$$
and thus such an operator would act in $\Ha_\beta$ as zero. Therefore 
an approximation of $\Theta$ by functions from $H_1$ would result in unitaries
that converge weakly to zero.

This situation can be visualized by the following 
\paragraph{Example} (3.1) \\
Consider the $H_1$-function $\,\Phi_{\delta,\ve}(x)$,
$$
\Phi_{\delta,\ve}(x) := \vp_\ve(x) - \vp_\ve(x + \delta) \in H_1, 
$$
with
$$
\vp_\ve(x) := \left\{\ba{cl} 1 & \mbox{ for } x \leq -\ve \\
-x/\ve & \mbox{ for } - \ve \leq x \leq 0 \\
0 & \mbox{ for } x \geq 0 \ea \right.
$$
as an approximation to the step function,
$$
\lim_{\delta \ra \infty \atop \ve \ra 0} \Phi_{\delta,\ve}(x) =
\Theta(x).
$$
Then
$$
\wt \Phi_{\delta,\ve}(p) = \frac{1 - e^{ip\ve}}{\ve p^2} (1 - e^{ip\delta})
$$
and
$$
\|\Phi_{\delta,\ve}\|^2_\beta = \int_{-\infty}^\infty \frac{dp}{2\pi} 
\frac{p}{1 - e^{-\beta p}}
| \wt \Phi(p)|^2 =
16 \int_{-\infty}^\infty \frac{dp}{2\pi} \frac{p}{1 - e^{-\beta p}}
\frac{\sin^2 p \ve/2}{\ve^2 p^4} \sin^2 p \delta/2 , 
$$
$$
\|\Phi\|^2_\beta \geq c \int_0^{1/\delta} dp \; \delta^2 = c \delta
$$
for $\beta/\delta$, $\ve/\delta \ll 1$ and $c$ a constant. Thus for
$\delta \ra \infty\,$, $\,\|\Phi_\delta\|_\beta \ra \infty$. Also $\,
\|\Phi_\delta - f\|_\beta \ra \infty$ since 
$$
\|\Phi_\delta - f\|_\beta \geq \|\Phi_\delta\|_\beta - \|f\|_\beta \ra \infty
\qquad \forall \; \|f\|_\beta < \infty
$$
and thus
$$
|\langle \Omega|e^{-ij_f} \; e^{ij_{\Phi_\delta}}|\Omega\rangle| =
e^{-\frac{1}{2} \|\Phi_\delta - f\|_\beta^2} \ra 0.
$$
But $e^{ij_f}|\Omega\rangle$, $\|f\|_\beta < \infty$, is total in 
$\Ha_\beta$ and thus $e^{ij_{\Phi_\delta}}|\Omega\rangle$ and therefore
$e^{ij_{\Phi_\delta}}$ goes weakly to zero. 

However the automorphism
$$
e^{ij_f} \ra e^{-ij_{\Phi_\delta}} e^{ij_f} e^{ij_{\Phi_\delta}} =
e^{i \sigma(\Phi_\delta,f)} e^{ij_f}
$$
converges since
$$
\sigma(f,\Phi_\delta) = - \frac{1}{2\pi\ve} \left( \int_{-\ve}^0 -
\int_{-\ve - \delta}^{-\delta}\right) dx \; f(x) 
\st{\delta \ra \infty}{\longrightarrow}
- \frac{1}{2\pi\ve} \int_{-\ve}^0 dx \; f(x)
\st{\ve\ra 0}{\longrightarrow}
-\frac{1}{2\pi}\,f(0).
$$
The divergence of $\|\Phi_{\delta,\ve}\|$ is related to the well-known
infrared problem of the massless scalar field in $(1\!+\!1)$ dimensions and
various remedies have been proposed \cite{S}. We take it as a sign
that one should enlarge $\,\A^l_c\,$ to some $\,\bar \A^l_c\,$ and work in the 
Hilbert space $\bar \Ha\,$ generated by $\,\bar \A^l_c\,$ on the natural extension 
of the state. Thus we add to $\,\A^l_c\,$ the idealized element 
$\,e^{i 2\pi j_{\vp_\ve}} = U_\pi\,$
and keep $\sigma$ and $\omega_\beta$ as before.
Equivalently we take the automorphism $\,\gamma\,$ generated by $\,U_\pi\,$ and
consider the crossed product $\,\bar \A^l_c = \A^l_c \st{\gamma}{\bowtie} 
\bf Z\,$. There is a natural extension $\,\bar \omega\,$ to $\,\bar \A^l_c\,$ 
and a natural isomorphism of $\,\bar \Ha\,$ and $\,\bar \A^l_c|\bar \Omega\rangle$.
Here $\bar \Ha$ is the countable orthogonal sum of sectors with $n$
particles created by $\,U_\pi$. Thus,
$$
\langle \Omega|e^{ij_f} U_\pi|\Omega\rangle = 0 \eqno(3.2)
$$
means that $U_\pi$ leads to the one-particle sector, 
in general
$$
\langle \Omega|U_\pi^{*n} \, e^{ij_f} U_\pi^m|\Omega\rangle =
\delta_{nm}\,\omega_\beta(\gamma^n\,e^{ij_f}).
$$ 
The quasifree automorphisms on $\,\A^l_c$ (e.g. $\tau_t$) 
can be naturally extended to $\,\bar \A^l_c\,\,$, $\tau_t \,U_\pi~=~
e^{i\pi j_{\vp_{\ve,t}}}$, $\,\vp_{\ve,t}(x) = \vp_\ve(x\!+\!t)\,$  and since
$\vp_\ve - \vp_{\ve,t} \in H_1$ $\,\forall \; t$, this does not lead out
of $\bar \A^l_c$. 

$U_\pi$ has some features of a fermionic field since
$$
\sigma(\vp_\ve,\tau_t \vp_\ve) = - \sigma(\vp_\ve,\tau_{-t} \vp_\ve) =
\frac{1}{4\pi}\left\{\ba{cl} 1 & \mbox{ for } t > \ve \\
\frac{2t}{\ve} - \frac{t^2}{\ve^2} & \mbox{ for } 0 \leq t \leq \ve \ea \right. . \eqno(3.3)
$$
More generally we could define $U_\alpha = e^{i \sqrt{2\pi\alpha} j_{\vp_\ve}}$
and get from (3.3) with 
$$
\sgn(t) = \Theta(t) - \Theta(-t) = \left\{
\begin{array}{cl} 1 & {\rm for} \qquad t > 0 \\
0 & {\rm for} \qquad t = 0 \\
-1 & {\rm for} \qquad t <  0.
\end{array} \right.
$$
\paragraph{Proposition} (3.4)
\beqan
U_\alpha \tau_t U_\alpha &=& \tau_t(U_\alpha) U_\alpha \,
e^{i\,\alpha\,\sgn(t)/2}, \\
U^*_\alpha \tau_t U_\alpha &=& \tau_t(U_\alpha) U^*_\alpha \,
e^{i\,\alpha\,\sgn(t)/2} \quad \forall \; |t| > \ve.
\eeqan

\paragraph{Remark} (3.5) \\
We note a striking difference between the general case of 
anyon statistics 
and the two particular cases --- Bose ($\alpha = 2\cdot 2n\pi$) or Fermi 
($\alpha = 2(2n+1)\pi$) 
statistics. Only in the latter two cases parity $P$ (2.12:4) is an
automorphism of the extended algebra generated through $U_\alpha$. Thus $P$
which was destroyed in $\A^l_c$ is now recovered for two subalgebras.\\[2pt]

The particle sectors are orthogonal in any case
$$
\langle \Omega| U^*{}^n_\alpha \,e^{ij_f} U^m_\alpha|\Omega\rangle =
0 \quad \forall \; n \neq m, \; f \in H_1.
$$
Furthermore, sectors with different statistics are orthogonal
$\langle\Omega\vert U^*_\alpha U_{\alpha'}\vert\Omega\rangle = 0, \, \alpha
\not= \alpha'$, thus if we adjoin $U_\alpha, \, \forall \alpha\in{\bf R_+}$,
$\,\bar\Ha_\beta$ becomes nonseparable.
 
Next we want to get rid of the ultraviolet cut-off and let $\ve$ go to
zero. Proceeding the same way we can extend $\sigma$ and $\tau_t$ but 
keeping $\omega$ the sectors abound. The reason is that
$\vp_\ve \st{\ve \ra 0}{\longrightarrow} \Theta(x)$ and
$$
\|\Theta - \Theta_t\|^2 = \int_{-\infty}^\infty 
\frac{dp \; p}{1 - e^{-\beta p}} \frac{|1 - e^{itp}|^2}{p^2}
$$
is finite near $p = 0$ but diverges logarithmically for $p \ra \infty$.
This means that $e^{ij_f} e^{ij_\Theta}|\Omega\rangle$, $f \in H_1$
gives a sector where one of these particles (fermions, bosons or anyons)
is at the point $x = 0$ and is orthogonal to $e^{ij_f} e^{ij_{\Theta_t}}
| \Omega\rangle\,$ $\forall \; t \not= 0$. Thus the total Hilbert space is not
separable and the shift $\tau_t$ is not even weakly continuous, so
there is no chance to make sense of $\frac{d}{dt} \tau_t e^{ij_\Theta}$.

So far, only one chiral component has been considered. When both chiralities
are present, no significant changes arise in the construction described. The
only point demanding for some care is the anticommutativity between left- and
right- moving fermions, which asks for an even larger extension of the current
algebra by extending its test--functions space. 

So, the Weyl algebra $\,\A_c = \A_c^r \otimes \A_c^l \,$ is now generated by the
unitaries
$$
W(f_r, f_l) = e^{i\int\left( f_r(x)j_r(x) + f_l(x)j_l(x)\right)dx},
$$
with $\sigma(f_l, g_l)$ given by (2.11) and $\sigma(f_r, g_r) = 
-\sigma(f_l, g_l)$. The minimal extension of $\,\A_c\,$ is then obtained by
adding two idealized elements, 
$$
U_\pi^l := W(c_l, \,2\pi(1-\vp_\ve)) \quad \mbox{and }\quad
U_\pi^r := W(2\pi\vp_\ve, \,c_r)
$$
$$ 
c_l - c_r = (2k + 1)\pi, \quad k \in {\bf Z} \qquad
(\mbox{e.g. } c_r = \pi/2 = -c_l)
$$
They generate for $\ve \ra 0$ (not inner) automorphisms of $\,\A_c\,$
 $$
 U_\pi^{r(l)} : \quad \gamma_{r(l)}\,W(f_r, f_l) = e^{if_{r(l)}(0)}
 \,e^{\frac{i}{4}\int_{-\infty}^\infty f'_{r(l)}(y)dy}\,W(f_r,f_l)
 $$
 in which for the two subalgebras $\,\{W(\bar f, \bar f)\}\,$ and $\,\{W(\bar f,
 -\bar f)\},\,\, \bar f \in \D_o\subset \C_o^\infty,\,$ the
 vector and axial gauge transformations can easily be traced back.
 
 In addition to the obvious replacement of (3.4),  also the following relation
 holds
 $$
 U_\pi^{r(*)}\, U_\pi^l = - U_\pi^l \, U_\pi^{r(*)}.
 $$
 
 Thus we can identify the chiral components of the fermion field with the so
 constructed unitaries
 $$
 \ba{ccl}
 \psi_r(x) & = & \dlim_{\ve\ra 0}\,\exp\,\{i2\pi\dint^{\infty}_{-\infty}
 \vp_\ve(y-x)j_r(x')dx' \pm i\frac{\pi}{2}\dint_{-\infty}^\infty
 j_l(x')dx'\}\\[8pt] 
 \psi_l(x) & = & \dlim_{\ve\ra 0}\,\exp\,\{\mp
 i\frac{\pi}{2}\dint_{-\infty}^\infty j_r(x')dx' 
 + i2\pi\dint^{\infty}_{-\infty} \vp_\ve(x-y) j_l(x')dx'\}
 \ea  \eqno(3.6)
 $$
 
 In general, we could define an extension of the algebra $\,\A_c\,$ through the
 abstract elements $\,U_\alpha^{r(l)}\,$
 \beqan
 U_{\alpha}^r &:=& W\left(\sqrt{2\pi\alpha}\, \vp_\ve, \,
 \frac{1}{2}\sqrt{\frac{\pi\alpha}{2}}\,\right) \\ 
 U_{\alpha}^l &:=& W\left(-\frac{1}{2}\sqrt{\frac{\pi\alpha}{2}}, 
 \,\sqrt{2\pi\alpha}\, (1-\vp_\ve)\right) \no
 \eeqan
 
 Propositon (3.4) extends also for the non-chiral model generalization.
 As expected, admitting arbitrary values for $\alpha$, we get a very rich field
 structure where definite statistic behaviour is preserved only within a given
 field class (fixed value of $\alpha$), so that even different fermions
 (with different ``2 x odd" values of $\alpha$) do not anticommute but instead 
 follow the general fractional statistics law.
\vspace{0.6cm}

\section{Anyon fields in $\pi_{\bar \omega}(\bar \A_c)''$}
Next we shall use another ultraviolet limit to construct local fields
which obey some anyon statistics. Of course quantities like
$$
[\Psi^*(x),\Psi(x')]_\alpha := 
\Psi^*(x) \Psi(x') e^{i \frac{2\pi-\alpha}{4} \sgn(x'-x)} 
+ \Psi(x') \Psi^*(x) e^{-i \frac{2\pi-\alpha}{4} \sgn(x'-x)} =
\delta_\alpha(x\!-\!x')
$$
will only be operator valued distributions and have to be smeared to give
operators. Furthermore in this limit the unitaries we used so far have to 
be renormalized so that the distribution $\delta_\alpha(x\!-\!x')$ (which
should be localized at a point and for certain values of $\alpha$ is
supposed to coincide with the ordinary $\delta$-function) gets a factor
1 in front. A candidate for $\Psi(x)$ will be 
$$
\Psi(x) := 
\lim_{\ve \ra 0} n_\alpha(\ve) \exp \left[ i \sqrt{2\pi\alpha}
\int_{-\infty}^\infty
dy \; \vp_\ve(x-y) j(y) \right] 
$$
with $\vp_\ve$ from (3.1) and $n_\alpha(\ve)$ a suitably chosen
normalization.
With the shorthand $\vp_{\ve,x}(y) = \vp_\ve(x-y)$ and $\tilde\alpha =
\sqrt{2\pi\alpha}$ we can write
\beqan
\Psi^*_\ve(x) \Psi_\ve(x') &=& 
\exp \left\{ i \,2\pi\alpha \,\sigma(\vp_{\ve,x},\vp_{\ve,x'}) \right\}
\exp \left\{ i \tilde\alpha \, j_{\vp_{\ve,x'} - \vp_{\ve,x}}\right\}, \\
\Psi_\ve(x') \Psi^*_\ve(x) &=&
\exp \left\{- i \,2\pi\alpha \,\sigma(\vp_{\ve,x},\vp_{\ve,x'}) \right\}
\exp \left\{ i \,\tilde\alpha \,j_{\vp_{\ve,x'} - \vp_{\ve,x}}\right\}.
\eeqan
We had in (3.3)
$$
4\pi\sigma(\vp_{\ve,x},\vp_{\ve,x'}) = \sgn(x-x')
\left\{ \Theta(|x-x'| - \ve) + \Theta(\ve - |x-x'|)
\frac{(x - x')^2}{\ve^2} \right\} 
$$
$$
=: \sgn(x-x') D_\ve(x-x')
$$
and thus
$$
[\Psi^*_\ve(x),\Psi_\ve(x')]_\alpha = 
2n_\alpha(\ve)^2 \cos \left[\sgn(x-x')
\left( \frac{\pi}{2} - \frac{\alpha}{4}(1 - D_\ve(x-x'))\right)\right]
\exp \left\{ i \tilde\alpha j_{\vp_{\ve,x'} - \vp_{\ve,x}}\right\}.
$$
Note that for $|x-x'| \geq \ve\,$ the argument of the $\cos$ becomes 
$ \pm \pi/2$, so the $\alpha$--commutator vanishes, in agreement with (3.4). 
To manufacture a $\delta$-function for $|x-x'|
\leq \ve$ we note that $\cos(...) > 0$ and $\omega_\beta(e^{i\alpha j}) >
0$, 
so we have to choose $n_\alpha(\ve)$ such that
$$
2 n_\alpha^2(\ve)\ve \int_{-1}^1 d\delta \cos \left( \frac{\pi}{2} -
\frac{\alpha}{4}(1 - \delta^2) \right) \cdot \omega_\beta
\left(\exp \left\{ i \tilde\alpha j_{\vp_{\ve,x-\ve\delta} -
\vp_{\ve,x}}\right\}\right)
= 1
$$
and to verify that for $\ve \ra 0_+$ $\,[.,.]_\alpha\,$ converges  
strongly to a $c$-number. For the latter to be finite we have to smear
$\Psi(x)$ with $L^2$-functions $g$ and $h$:
$$
\int dx dx' g^*(x) h(x') [\Psi^*_{\ve}(x),\Psi_{\ve}(x')]_\alpha = 
\int dx dx' g^*(x) h(x') 2n_\alpha(\ve)^2 \cos(\;)
\exp \left\{ i \tilde\alpha j_{\vp_{\ve,x'} - \vp_{\ve,x}}\right\}.
$$
This converges strongly to $\langle g|h\rangle$ if for $\ve \ra 0_+$
$$
\left\langle\exp \left\{- i \tilde\alpha j_{\vp_{\ve,x'} -
\vp_{\ve,x}}\right\}
\exp \left\{ i \tilde\alpha j_{\vp_{\ve,y'} -
\vp_{\ve,y}}\right\}\right\rangle_\beta
- \left\langle\exp \left\{- i \tilde\alpha j_{\vp_{\ve,x'} -
\vp_{\ve,x}}\right\}
\right\rangle_\beta\, \left\langle
\exp \left\{ i \tilde\alpha j_{\vp_{\ve,y'} -
\vp_{\ve,y}}\right\}\right\rangle_\beta
\ra 0
$$
for almost all $x,x',y,y'$. Now
$$
\langle e^{-ij_a} e^{ij_b} \rangle_\beta = \langle e^{-ij_a}\rangle_\beta
\langle e^{ij_b}\rangle_\beta
\exp \left[\int_{-\infty}^\infty \frac{dp \;p}{1 - e^{\beta p}}
\wt a(-p) \wt b(p) \right].
$$
In our case this last factor is
\beqan
\lefteqn{ \int_{-\infty}^\infty \frac{dp \;p}{1 - e^{-\beta p}}
\frac{|1 - e^{ip\ve}|^2}{\ve^2 p^4} (e^{ipx} - e^{ipx'})
(e^{-ipy} - e^{-ipy'}) = } \\
&=& \int_{-\infty}^\infty \frac{dp\;2(1-\cos p)}{p^3(1-e^{-\beta p/\ve})}
(e^{ipx/\ve} - e^{ipx'/\ve})(e^{-ipy/\ve} - e^{-ipy'/\ve}).
\eeqan
For fixed $\beta \neq 0$ and almost all $x,x',y,y'$ this converges to zero
for $\ve \ra 0$ by Riemann-Lebesgue. 
Thus a $\delta$-type distribution is recognized, however the particular
structure of the singularity we shall discuss later on, already on the
basis of the state.

In the same way one sees that
$\exp \left\{i \tilde\alpha j_{\vp_{\ve,x} + \vp_{\ve,x'}} \right\}$
converges strongly to zero and that the $\Psi_{\ve,g}$ are a strong Cauchy
sequence for $\ve \ra 0$. To summarize we state
\paragraph{Theorem} (4.1) \\
$\Psi_{\ve,g}$ converges strongly for $\ve \ra 0$ to an operator
$\Psi_g$ which for $\alpha = 2\pi$ satisfies 
$$
[\Psi_g^*,\Psi_h]_+ = \langle g|h\rangle, \qquad
[\Psi_g,\Psi_h]_+ = 0.
$$
If $supp\, g < supp\, h$, 
$$
\Psi_g^* \Psi_h \, e^{i\frac{2\pi - \alpha}{4}} +
\Psi_h \Psi_g^* \, e^{-i\frac{2\pi - \alpha}{4}} = 0 \qquad \forall \alpha.
$$

Furthermore we have to verify the claim (1.5) that also for $\Psi_g$ the
current $j_f$ induces the local gauge transformation $g(x) \ra e^{2i\alpha f(x)}
g(x)$. For finite $\ve$ we have 
$$
e^{ij_f} \Psi_{\ve,g} e^{-ij_f} = 
\Psi_{\ve, e^{i 2\pi\alpha \sigma(f,\vp_\ve)}g}
$$
and for $\ve \ra 0_+$ we get
$\sigma(f, \vp_\ve) \ra \frac{1}{2\pi}f(0)\,$, so that $\,\sigma(f,
\tau_x\vp_\ve) = \frac{1}{2\pi}f(x)$. \\[2pt]

To conclude we investigate the status of the ``Urgleichung'' in our
construction. It is clear that the product of operator valued distributions on
the r.h.s. can assume a meaning only by a definite limiting prescription.
Formally it would be
$$
\Psi(x)\Psi^*(x)\Psi(x) = [\Psi(x), \Psi^*(x)]_+\Psi(x) - 
\Psi^*(x)\Psi(x)^2 = \delta(0)\Psi(x) - 0.
$$
From (2.12:5) we know
$$
\frac{1}{i} \frac{\partial}{\partial x} \Psi_\ve(x) = 
\frac{\sqrt{2\pi\alpha}}{2} \,[\bar \jmath(x),\Psi_\ve(x)]_+, \qquad
\bar \jmath(x) = \frac{1}{\ve} \int_{x-\ve}^x dy \; j(y).
$$
Using 
$$
j_{\vp'} e^{ij_\phi} = \left. \frac{1}{i}\frac{\partial}{\partial\alpha}
e^{i\frac{\alpha}{2}\sigma(\vp',\vp)} e^{ij_{\vp+\alpha
\vp'}}\right\vert_{\alpha = 0}
$$ 
one can verify that the limit $\ve \ra 0_+$ exists for the expectation
value with a total set of vectors and thus gives densely defined (not
closable) quadratic forms. They do not lead to operators but we know from
(2.7), (2.8) that they define operator valued distributions for test
functions from $H_1$. Thus one could say that in the sense of operator
valued distributions the Urgleichung holds
$$
\frac{1}{i} \frac{\partial}{\partial x} \Psi(x) = \frac{\sqrt{2\pi\alpha}}{2}
\,[j(x),\Psi(x)]_+ . \eqno(4.2)
$$

The remarkable point is that the coupling constant $\lambda$ in
(1.1) is related to the statistics parameter $\alpha$. For fermions one
has a solution only for $\lambda = \sqrt{2\pi}$. Of course one could for
any $\lambda$ enforce fermi statistics by renormalizing the bare
fermion field $\psi \ra \sqrt{Z}\;\psi$, $j \ra Zj$ with a suitable
$Z(\lambda)$ but this just means pushing factors around. Alternatively
one could extend $\A_c$ by adding $e^{i\sqrt{2\pi\alpha}\,j_{\vp_\ve}}$, for all
$\alpha \in {\bf R_+}$. Then one gets in $\Ha_\omega$ uncountably many
orthogonal sectors, one for each $\alpha$, and in each sector a different
Urgleichung holds. Thus different anyons live in 
orthogonal Hilbert spaces and $e^{i\sqrt{2\pi\alpha}\,j_{\vp_\ve}}$ is not even 
weakly continuous in $\alpha$. If $\alpha$ is tied to $\lambda$ it is clear that
an expansion in $\lambda$ is doomed to failure and will never reveal
the true structure of the theory.
\vspace{0.6cm}

\section{Correlation functions in a KMS-state}

In terms of distributions, the  $\tau$-KMS-state $\omega_\beta$ over
$\A^l\,$ reads
\beqa
\langle \psi^*(x)\psi(x')\rangle_\beta &=&
\lint_{-\infty}^\infty dp \frac{e^{-ip(x-x')}}{2\pi\,(1+e^{\beta p})} \\
\no
&=& 
-\frac{1}{2\pi}\sum_{n=-\infty}^{\infty}\frac{(-1)^n}{i(x-x'+i\ve)-n\beta}
\eeqa

It can be also represented in a form that makes the thermal
contributions explicit
$$
\langle \psi^*(x)\psi(x')\rangle_\beta =  
\frac{i(x-x')}{\pi}\sum_{n=1}^\infty \frac{(-1)^n}{(x-x')^2 + n^2\beta^2}
- \frac{1}{2\pi[i(x-x')-\ve]}
$$
or, for a better comparison with the dressed-fermion correlators, as
\beq 
\langle \psi^*(x)\psi(x')\rangle_\beta =
\frac{i}{2\beta\,\sinh{\frac{\pi(x-x'+i\ve)}{\beta}}}\, ,
\eeq
respectively
\beq
\langle \psi(x')\psi^*(x)\rangle_\beta =
-\frac{i}{2\beta\,\sinh{\frac{\pi(x-x'-i\ve)}{\beta}}}\, .
\eeq

The anyon fields that are present in the v. Neumann algebra
$\bar\pi_\beta(\bar\A_c)''$, associated to the extended current algebra
$\bar\A_c$, are of the form 
\beq
\Psi_\alpha(x) \sim
\exp \left[ i \sqrt{2\pi\alpha} 
\int_{-\infty}^\infty dy \; f^x(y) j(y) \right]\, 
\eeq
with an appropriately chosen smearing function $f$ (see Sect. 3, 4;
$f^x(y) = f(x-y)$).

The fields (5.4) are elements of a Weyl algebra, so they obey the usual
multiplication law
$$
W(f)W(g) = W(f+g)\,e^{-i\sigma(f,g)/2}\, ,
$$
with $\sigma$ --- the symplectic form of the algebra.
Recalling the symplectic form inherited by $\,\bar\pi_\beta(\bar\A_c)''\,$
from the current algebra $\,\A_c$ (so, actually from
$\,\pi_\beta(\A_c)''$), (2.11), and with (2.12,3) in mind, the correlation
functions of interest can be easily obtained. 

We start with the relation
\beq
\omega (e^{ij_f}e^{ij_g}) = \exp\left\{-\frac{1}{2}\left(\omega(j_f^2) +
\omega(j_g^2) + 2\omega(j_fj_g)\right)\right\} 
\eeq
with
$$
\omega_\beta(j_fj_g) = 
- \int\frac{dydy'f(y)g(y')}{(2\beta)^2\,\sinh^2
\frac{\pi(y-y'+i\ve)}{\beta}}\, , \qquad \ve \ra 0_+ \, .
$$ 
The shift $\tau_t: \, f(y) \ra f(y-t)$ can be continued analytically for
$f$ in the upper strip $0 < \mbox{Im } t < \beta$ and for $g$ in the lower
strip $0 > \mbox{Im } t > -\beta$. This reflects the KMS-property of
$\omega_\beta$ and
also serves as an ultraviolet cut-off if we consider $\ve$ as arising
from $\tau_{i\ve}j_f$. We are interested in $f^x(y) \sim -\Theta(x-y)$, 
$\,g^{x'}(y') \sim \Theta(x'-y')$ and then an infrared cut-off $\Theta(-y)
\ra \Theta(-y-\delta)$ will cancel out but to get rid of the ultraviolet
cut-off we have to define
\beq
\Psi(x) =
\lim_{\ve \ra 0} 
n_\alpha(\ve) \exp \left\{ i \sqrt{2\pi\alpha}
j_{\Theta_\ve^x} \right\}\, =: \Psi_\alpha(x)\, .
\eeq

In the expectation values in (5.5) the limits of the integrals are
$\int_{-\delta}^x \int_{-\delta}^{x'}\,$,
$\,\int_{-\delta}^x\int_{-\delta}^x\,$ and
$\,\int_{-\delta}^{x'}\int_{-\delta}^{x'}\,$ for $\,j_fj_g, \,\, j_f^2$
and $\,j_g^2\,$ respectively. We only have to work out the first, the
others are special cases, and we can scale $\pi/\beta$ away:
$$
\int_{-\delta}^x dy \int_{-\delta}^{x'}dy' \sinh^{-2}(y-y'+i\ve) = 
\ln\frac{\sinh(x-x'+i\ve)}{\sinh(x+\delta+i\ve)} - 
\ln\frac{\sinh(-\delta-x'+i\ve)}{\sinh (i\ve)}\, .
$$

To evaluate (5.5), we have to subtract from this expression $1/2$ the
same with $x'=x$ and $1/2$ the same with $x=x'$. Then for $\delta
\ra\infty$ the $\delta$-terms cancel out but what remains (already with
the proper coefficient) is 
$$
(\alpha/2\pi)\left[-\ln \sinh{\frac{\pi(x-x'+i\ve)}{\beta}}+\ln
(i\ve)\right]\, . 
$$
Upon exponentiation we get for $\alpha = 2\pi$ the fermion KMS two-point
function and a factor $\ve$ which has to be compensated by the
renormalization of $\Psi$, that is for fermions the renormalization
parameter should be $n_{2\pi}(\ve) = (2\pi\ve)^{-1/2}$ and for the general
case of an arbitrary $\alpha\,$: $\,n_\alpha(\ve) = 
(2\pi\ve)^{-\alpha/4\pi}$.

For all $\alpha$'s the two-point function (for $x \not= x'$ and $\beta =
\pi$)
\beq
\langle \Psi_\alpha^*(x) \Psi_\alpha(x') \rangle_\beta = 
\langle \Psi_\alpha(x) \Psi_\alpha^*(x') \rangle_\beta =
\left(\frac{i}{2\beta \sinh(x-x')}\right)^{\alpha/2\pi} =: S_\alpha(x-x')
\eeq
has the desired properties 
\begin{enumerate}
\item \Un{Hermiticity}: 
$$
S_\alpha^*(x) = S_\alpha(-x) \, \Longleftrightarrow \, \langle
\Psi_\alpha^*(x)\Psi_\alpha(x')\rangle_\beta^* = \langle
\Psi_\alpha^*(x')\Psi_\alpha(x)\rangle_\beta\,; 
$$
\item \Un{$\alpha$-commutativity}: 
$$
S_\alpha(-x) = e^{i\alpha/2}S_\alpha(x) \, \Longleftrightarrow \,  
\langle\Psi_\alpha(x')\Psi_\alpha^*(x)\rangle_\beta =
e^{i\alpha/2}\langle\Psi_\alpha^*(x)\Psi_\alpha(x')\rangle_\beta \, ;
$$
\item \Un{KMS-property}:
$$
S_\alpha(x) = S_\alpha(-x+i\pi) \, \Longleftrightarrow \, 
\langle\Psi_\alpha^*(x)\Psi_\alpha(x')\rangle_\beta = 
\langle\Psi_\alpha(x')\Psi_\alpha^*(x+i\pi)\rangle_\beta\, .
$$
\end{enumerate}

\noindent
{\bf Remark}

\noindent
If the deformation parameter $\exp{i\alpha/2} = q \not\in {\bf S}^1$ but
instead $q \in (-1,1)$ then there are no KMS-states since there do not
exist even translation invariant states \cite{HN}. 

\vspace{10pt}
For $\alpha = 2\pi$ Eq.(5.7) reads
\beq
\langle\Psi_{2\pi}^*(x)\Psi_{2\pi}(x')\rangle_\beta = \lim_{\ve\ra
0_+}\frac{i}{2\beta\,\sinh \frac{\pi(x-x'+i\ve)}{\beta}}\, .
\eeq

For $f(y)\ra\Theta(x-y)$, $\,g(y')\ra -\Theta(x'-y')$ nothing changes so
we verify the realtion (5.2) $\,\leftrightarrow\,$ (5.3),
$$
\langle\Psi_{2\pi}^*(x)\Psi_{2\pi}(x')\rangle_\beta =
\langle\Psi_{2\pi}(x)\Psi_{2\pi}^*(x')\rangle_\beta\, .
$$
 
For $\alpha = 4\pi$ we get like for the $j$'s
\beq
\langle\Psi_{4\pi}^*(x)\Psi_{4\pi}(x')\rangle_\beta =
-\frac{1}{\left(2\beta\,\sinh{\frac{\pi(x-x'+i\ve)}{\beta}}\right)^2}\, , 
\eeq
whereas for $\alpha = 6\pi$ we get a different kind of fermions
\beq
\langle\Psi_{6\pi}^*(x)\Psi_{6\pi}(x')\rangle_\beta =
-\frac{i}{\left(2\beta\,\sinh{\frac{\pi(x-x'+i\ve)}{\beta}}\right)^3}\, .
\eeq
They are not canonical fields since
$$
-[\Psi_{6\pi}^*(x),\Psi_{6\pi}(x')]_\alpha 
= [\Psi_{6\pi}^*(x),\Psi_{6\pi}(x')]_+ = 
-\frac{1}{8\pi^2}\left(\delta''(x-x') -
\frac{\pi^2}{\beta^2}\delta(x-x')\right)\,.
$$
This shows that local anticommutativity alone does not guarantee the
uniqueness of the KMS-state, one needs in addition the CAR-relations. The 
$\Psi_\alpha$'s, $\,\alpha \in (2{\bf N}+1)\pi\,,\,$ describe an infinity
of inequivalent fermions.

For the thermal expectation of the $\alpha$-commutator one finds
\beqa
\omega_\beta([\Psi^*(x), \Psi(x')]_\alpha) &=&
-i\left(-\frac{1}{2\beta\,\sinh{\frac{\pi(x-x'+i\ve)}{\beta}}}\right) 
^{\alpha/2\pi}
\no \\
&+&
i\left(-\frac{1}{2\beta\,\sinh{\frac{\pi(x-x'-i\ve)}{\beta}}}\right) 
^{\alpha/2\pi}
\eeqa
and for $\ve \ra 0$ a distribution is obtained where only the leading
singularity is temperature independant, namely
$$
\lim_{\ve\ra 0}
\left\{\left(\frac{1}{x-x'-i\ve}\right)^{\alpha/2\pi} 
- \left(\frac{1}{x-x'+i\ve}\right)^{\alpha/2\pi}\right\} = 
\, ? 
$$
We shall not attempt to make sense of the ``canonical" anyon commutators.
Another interesting topic --- the higher correlation functions, since the
limiting procedures involved might influence their properties, we shall
also discuss separately \cite{thc}.

\vspace{0.6cm}
\section{Internal symmetries}
We shall briefly discuss what happens in the the case of a fermion multiplet.
Then,
$$
\{ \psi_i^*(x), \psi_k(y)\} = \delta_{ik}\,\delta(x-y),
\qquad i = 1,2,...,N
$$
The CAR-algebra so defined posesses an obvious $U(N)$ symmetry
$$
\psi \longrightarrow U\,\psi, \quad U \in U(N)
$$
In analogy with the previous case we construct quadratic forms $\,j_k(x) =
\psi^*_k(x) \psi_k(x)\,$ which satisfy anomalous commutation relations:
\beq
[j_k(x), j_m(y)] = -\frac{i}{2\pi}\delta_{km}\delta'(x-y)
\eeq
and give rise to operators $\, j_{f_k} = \int j_k(x)f_k(x)dx, \,\, f \in H_1$.
 
Denote the corresponding Weyl operators with $W(f_1,...,f_N)$
$$
W(f_1,...,f_N) := \exp\{i\sum_{n = 1}^N j_{f_n}\}
$$
The current algebra (6.1) has a (global) $O(N)$ symmetry, 
\beq
j \rightarrow  Mj = j', \qquad M \in O(N)
\eeq

Genuine anticommuting fields can now be identified 
in an extension of the current algebra quite similar to the one for the
non-chiral one-colour case. More precisely, we have to allow for the 
existence in the new algebra of the following elements 
$$
U_{\pi_k} = e^{i2\pi\lint_{-\infty}^\infty \vp(x-x')j_k(x')dx' + 
i \dsum_{n=1\,, n \not= k}^N c_n \lint_{-\infty}^\infty j_n(x')dx'} =:
\psi_k(x), 
$$
\beq
c_k \in {\bf R}, \quad  c_k - c_n = (2l+1)\pi, \quad 
l\in{\bf Z}, \,\,\forall k,n\, . 
\eeq

For the elements so defined the following relation holds
$$
\psi^{*}_k(x) \psi_l(y) = \psi_l(y)\psi^{*}_k(x)
e^{-i\pi\,\delta_{kl}\,\sgn(y-x)}\,e^{-i(1-\delta_{kl})(c_k-c_l)}
$$
which would then lead (after an appropriate renormalisation) to the desired 
CAR's. 

How should one consider the element, obtained through the same ansatz, but
after a transformation (6.2) of the currents, i.e.
\beq
\psi'_k(x) =  e^{i2\pi\int_{-\infty}^x  M_{kl}j_l(x')dx' + (...)}
\eeq
Has it something to do with the $U(N)$-transformed fermion $\psi'_k$, i.e.
$$
e^{i2\pi\int_{-\infty}^x  M_{kl}j_l(x')dx'} \, \st{?}{\longleftrightarrow} \,
U_{kl}\psi_l(x)
$$

What one notices is that the $O(N)$-transformation, being (of course) an
isomorphism of the algebra $\,\bar \A_c\,$, is no longer an automorphism
of the total algebra. 

In such a symmetry nonpreservation a major limitation of the
usual bosonization scheme (with non-local fermi fields,
\cite{SC, SM}) has been recognized, e.g. the $O(N)$ symmetry of
a free scalar theory with $N$ fields, which emerges upon
bosonization of a theory with $N$ free Dirac fields with $U(N)\times U(N)$ 
chiral invariance, does not correspond to any subgroup of the fermion
symmetry group \cite{EW}. The bosonization scheme proposed therein 
preserves all symmetries through the passage from one formulation to the
other. However, this does not concern the complete cycle where, as we have
seen, symmetries may be (actually, necessarily are) spontaneously
destroyed.

We are faced with the similar situation also for the $U$-invariance we
mentioned at the beginning. Here, e.g. for $\,U(1)\,$, so
$$
\psi_k(x) \rightarrow e^{i\alpha}\psi_k(x)
$$
we get 
$$
\psi'_k(x) =  e^{i\alpha}e^{i2\pi\int_{-\infty}^x j_k(x')dx'} =
e^{i2\pi\int_{-\infty}^x j'_k(x')dx'}
$$
with
$$
j'_k(x') = j_k(x') + \frac{1}{2\pi}\bar\alpha(x'), \quad
\bar\alpha_{(k)} : \int_{-\infty}^x \bar\alpha(x') = \alpha(x)
$$
In the local case this still remains an automorphism of the extended algebra,
which is in agreement with the very idea of the construction presented, while
for a global transformation this is no longer the case, so for the total
algebra so constructed there is no global $U(1)$ symmetry present.

However, on the passage from the CAR-algebra $\,\A\,$ to the current algebra 
$\,\A_c\,$ contained in the v.Neumann algebra $\,\pi_\beta(\A)''$ the parity
has been broken, but also as an isomorphism of $\,\A_c\,$. 

Thus we are in a situation in which a particular (physically motivated)
extension $\,\bar\A_c,$ of the algebra of observables (the current algebra 
$\A_c\,$) is constructed, $\,\bar\A_c \supset \A_c$, such that 
$\,\not\exists \beta \in \mbox{Aut } \bar\A_c: \,\, \beta\vert_\A = \alpha \,$ 
for some $\, \alpha \in \mbox{Aut }\A_c\,$. This phenomenon we call 
{\it symmetry destruction}. It is related to the spontaneous collapse of a 
symmetry, discussed by Buchholz and Ojima in the context of supersymmetry 
\cite{BO} and is seen to be a field effect since in contrast to the spontaneous 
symmetry breaking \cite{NT}, it cannot occur in a finite-dimensional Hilbert 
space. 

\vspace{0.6cm}

\section{Concluding remarks}
To summarize we gave a precise meaning to eq.(1.2a,b,c) by starting with bare
fermions, $\A = {\rm CAR}(\bf R)$. The shift $\tau_t$ is an automorphism of $\A$
which has KMS-states $\omega_\beta$ and associated representations
$\pi_\beta$. In $\pi_\beta(\A)''$ one finds bosonic modes $\A_c$ with an
algebraic structure independent on $\beta$. Taking the crossed product with an
outer automorphism of $\A_c$ or equivalently augmenting $\A_c$ by an unitary
operator to $\bar\A_c$ we discover in $\bar \pi_\beta(\bar\A_c)''$ 
anyonic modes which satisfy the Urgleichung in a distributional sense.
For special values of $\lambda$ they are dressed fermions distinct from
the bare ones. From the algebraic inclusions CAR({\it bare}) $\subset
\pi_\beta(\A)'' \supset\A_c \subset \bar \A_c \subset  
\bar\pi_\beta(\bar\A_c)'' \supset$ CAR({\it dressed}) one concludes
that in our model it cannot be decided whether fermions or bosons are
more fundamental. One can construct the dressed fermions either from
bare fermions or directly from the current algebra.
Also the correlation functions in the temperature state in both cases
coincide that is in agreement with the uniqueness of the $\tau$-KMS
state over a CAR algebra ($\tau$ being the shift automorphism). The
corresponding anyonic two-point function evaluated for the special case
of a scalar field reproduces the recent result due to Borchers and Yngvason.

\vspace{0.6cm}

\section*{Acknowledgements}
The authors are grateful to H. Narnhofer, R. Haag and J. Yngvason for
helpful discussions on the subject of this paper.\\[2pt]

N.I. acknowledges the financial support from the ``Fonds zur F\"orderung der
wissenschaftlichen Forschung in \"Osterreich" under grant P11287--PHY and the
hospitality at the Institute for Theoretical Physics of University of Vienna.
\vspace{1cm}

\appendix
\newcounter{zahler}
\renewcommand{\thesection}{Appendix}
\renewcommand{\theequation}{\Alph{zahler}.\arabic{equation}}
\setcounter{zahler}{1}
\section{The field algebra as a crossed product}

The idea that the crossed product $C^\ast$-algebra extension is the tool
that makes possible construction of fermions (so, unobservable fields) 
from the observable algebra has been first stated in \cite{DHR}. There,
the problem of obtaining different field groups has been shown to amount
to construction of extensions of the observable algebra by the group
duals. Explicitly, crossed products of $C^\ast$-algebras by semigroups of
endomorphisms have been introduced when proving the existence of a compact
global gauge group in particle physics given only the local observables
\cite{DR}. Also in the structural analysis of the symmetries in the
algebraic QFT \cite{H} extendibility of automorphisms from a unital
$C^\ast$-algebra to its crossed product by a compact group dual becomes
of importance since it provides an analysis of the symmetry breaking
\cite{NT} and in the case of a broken symmetry allows for concrete
conlusions for the vacuum degeneracy \cite{BDLR}. 

The reason why a relatively complicated object --- crossed product over a
specially directed symmetric monoidal subcategory $\rm End \,\A$ of unital
endomorphisms of the observable algebra $\A$, is involved in
considerations in \cite{BDLR} is that in general, non-Abelian gauge
groups are envisaged. For the Abelian group $U(1)$ a significant
simplification is possible since its dual is also a group --- the group
{\bf Z}. On the other hand, even in this simple case the problem of
describing the local gauge transformations remains open. Therefore in the
Abelian case consideration of crossed products over a discrete group
offers both a realistic framework and reasonable simplification for the
analysis of the resulting field algebra. We shall briefly outline the
general construction for this case, for more details see \cite{IN}. 

We start with the CCR algebra $\A(\V_0,\sigma)$ over the real symplectic
space $\V_0$ with symplectic form $\sigma$, Eq.(2.11), generated by the
unitaries $\,W(f), \,f \in \V_0$ with
$$
W(f_1)W(f_2) = e^{-i\sigma(f_1,f_2)/2}W(f_1+f_2), \qquad 
W(f)^\ast = W(-f) = W (f)^{-1}.  
$$
Instead of the canonical extension $\,\bar\A(\V,\bar\sigma), \,
\V_0\subset\V\,$ \cite{AMS}, we want to construct another algebra $\F$,
such that ${\rm CCR}(\V_0)\subset\F\subset{\rm CCR}(\V)$ and we choose
$\,\V_0 = \C_0^{\,\infty}, \, \V = \partial^{-1}\C_0^{\,\infty}\,$. Any
free (not inner) automorphism $\alpha, \, \alpha\in{\rm Aut}\,\A\,$
defines a crossed product $\, \F = \A\, \st{\alpha}{\bowtie}\,{\bf Z}\,$. 
This may be thought as (see \cite{A})  adding to the initial algebra $\A$
a single unitary operator $U$ together with all its powers, so that one
can formally write $ \,\F = \sum_n \A\,U^n\,$, with $U$ implementing the
automorphism $\alpha$ in $\A, \,\, \alpha A = U\,A\,U^\ast, \,\forall
A\in\A$. Operator $\,U\,$ should be thought of as a charge-creating
operator and $\F$ is the minimal extension --- an important point in
comparison to the canonical extension which we find superfluous especially
when questions about statistical behaviour and time evolution are to be
discussed. With the choice 
\beq
\alpha W(f) = e^{i\sigma(\bar g,f)}W(f), \qquad \bar g\in\V\backslash\V_0,
\qquad \V_0\subset\V 
\eeq 
and identifying $U = W(\bar g)\,$, $\,\F\,$ is in a natural way a
subalgebra of CCR$(\V)$.

If we take for $\A$ the current algebra $\A_c$ and for $U$ --- the
idealized element $U_\pi$ to be added to it, we find an obvious
correspondence between the functional picture from Sec.3 and the crossed
product construction.  However, in the latter there is an additional
structure present which makes it in some cases favourable. Writing an
element $F \in \F$ as $\, F = \sum_n A_n U^n, \quad A_n \in \A\,$, we see
that it is convenient to consider $\F$ as an infinite vector space with
$U^n$ as its basic unit vectors and $A_n =:  (F)_n$ as components of $F$.
The algebraic structure of $\F$ implies that multiplication in this space
is not componentwise but instead 
$$ 
(F.G)_m = \dsum_n F_n\,\alpha^n\,G_{m-n}.  
$$

Given a quasifree automorphism $\rho \in {\rm Aut}\,\A$, it can be
extended to $\F$ if and only if the related automorphism $\gamma_\rho =
\rho\alpha\rho^{-1}\alpha^{-1}$ is inner for $\A$. Since $\gamma_\rho$ is
implemented by $W(\bar g_\rho - \bar g)$, this is nothing else but
demanding that $\bar g_\rho - \bar g \in \V_0$ and this is exactly the
same requirement as in the functional picture. This appears to be the case
for the space translations and also for the time evolution, but in the
absence of long-range forces \cite{IN}. 

Also a state $\omega(.)$ over $\A$ together with the representation
$\,\pi_\omega\,$ associated with it through the GNS-construction can be
extended to $\F$. The representation space of $\F$ can be regarded as a
direct sum of charge-$n$ subspaces, each of them being associated with a
state $ \omega\circ\alpha^{-n} $ and with $ \Ha_0 $, the representation
space of $ \A $, naturally imbedded in it. Since $\omega$ is irreducible
and $ \omega\circ\alpha^{-n} $ not normal with respect to it, the
extension of the state over $\A $ to a state over $\F$ is uniquely
determined by the expectation value with $\vert\Omega_0\rangle =
\vert\omega\rangle$ in this representation 
$$ 
\langle\Omega_k\vert
W^\ast(f) W(h) W(f)\vert\Omega_n\rangle = \delta_{kn} \, e^{-i\sigma
(f+n\bar g, h)}\omega(W(h))  
$$ 
where $ U_k\vert\Omega\rangle :=
\vert\Omega_k\rangle, \, \langle\Omega_k\vert\Omega_n\rangle =
\delta_{kn}\,$. This states nothing but orthogonality of the different
charge sectors, the same as in the functional description, Eq.(3.2). 

In the crossed product gauge automorphism is naturally defined with 
\beq
\gamma_\nu \,U^n = e^{2\pi i\nu n} U^n, \qquad \gamma_\nu\, W(f) = W(f) 
\,. 
\eeq 
Thus for the representation $\pi_\Omega$ one finds 
$$
\gamma_\nu\left(\vert F(f)^{(k)}\rangle\right) =
\gamma_\nu\left(W(f)\vert\Omega_k\rangle\right) = e^{2\pi i\nu
k}W(f)\vert\Omega_k\rangle, 
$$ 
that justifies interpretation of the vectors $\,\vert F(f)^{(k)}\rangle\,$
as belonging to the charge-$k$ subspace. However, $\A$ is a subalgebra of
$\F$ for the gauge group $\T = [0,1)$, while it is a subalgebra of CAR for
the gauge group $\T\otimes\bf R$. Thus the crossed product algebra so
constructed, being really a Fermi algebra, does not coincide with CAR but
is only contained in it.  In other words, such a type of extension does
not allow incorporation also of local gauge transformations which are of
main importance in QFT. 

Therefore we need a generalization of the construction in \cite{IN} which
would describe also the local gauge transformations. The most natural
candidate for a structural automorphism would be 
\beq 
\alpha_{\bar g_x}W(f) = e^{i\sum_{n=0}^{K}f^{(n)}(x)}W(f).  
\eeq 
However, it turns out that only for $n = 0$ the crossed product algebra so
obtained allows for extension of space translations as an automorphism of
$\A$ --- the minimal requirement one should be able to meet. Already first
derivative gives for the zero Fourier component of the difference $\bar
g_{x_\delta} - \bar g_x$ an expression of the type $\int y^{-1}\delta(y)
dy$, so it drops out of $\,\C_0^{\,\infty}$. So, the automorphism of
interest reads 
\beq
\alpha_{\bar g_x} W(f) = e^{if(x)}W(f)  
\eeq 
and can be interpreted as being implemented by $\,W(\bar g_x)$ with $\bar
g_x = 2\pi\,\Theta (x\!-\!y)$.  Correspondingly, the operator we add to
$\A$ through the crossed product is 
\beq 
U_x = e^{i 2\pi\int_{-\infty}^{x}j(y) dy}.  
\eeq
Compared to \cite{IN} this means an enlargment of the test functions space
not with a kink but with its limit --- the sharp step function. In a
distributional sense it still can be considered as an element of
$\,\partial^{-1}\V_0\,$ for some $\,\V_0\,$ since the derivative of $\bar
g_x$ has bounded zero Fourier component. Similarly, the extendibility
condition for space translations is found to be satisfied, $\bar
g_{x_\delta}-\bar g_x \in\V_0$ so that in the crossed product shifts are
given by 
\beq 
\bar\tau_{x_\delta}U_x = V_{x_\delta}U_x, \qquad
V_{x_\delta} = W(\bar g_{x_\delta} - \bar g_x).  
\eeq 
Note that shifts do not commute with the structural automorphism
$\alpha_{\bar g_x}$, $\,\tau_{x_\delta}\alpha_{\bar g_x}W(f) \not=
\alpha_{\bar g_x}\tau_{x_\delta}W(f)$. Since 
\beq 
\sigma(\bar g_x, \bar g_{x_\delta}) = -\pi\sgn(\delta), 
\eeq 
already the elements of the first class are anticommuting and we identify
$U_x =:  \psi(x)$.  Then (A.4) (after smearing with a function from
$\C_0^{\,\infty}$) is nothing else but (1.5), i.e. the statement (or
requirement) that currents generate local gauge transformations of the
so--constructed field. Any scaling of the function which defines the
structural automorphism $\,\alpha_{\bar g_x}\,$ destroys relation (A.7) 
and fields obeying fractional statistics are obtained instead.  This is
effectively the same as adding to the algebra $\,\A\,$ the element
$U_\alpha$ with $\alpha = 2\pi\mu, \, \mu$ being the scaling parameter. 

However, the crossed product offers one more interesting possibility: when
for the symplectic form in question instead of (A.7) (or its direct
generalization $\,\sigma(\bar g_x, \bar g_{x_\delta}) = (2n + 1)\pi, n \in
\bf Z \,)$ another relation takes place, $\,\sigma(\bar g_x, \bar
g_{x_\delta}) = (2n+1)/\bar n^2\,$ for some fixed $\,\bar n\in \bf Z$, the
crossed product acquires a zone structure, with $\,2n\bar n$--classes
commuting, $\,(2n+1)$--classes anticommuting and elements in the classes
with numbers $m \in {\bf Z/Z}_{\bar n}$ obeying an anyon statistics with
parameter $\,r = \sqrt{2n + 1}\,m/\bar n\,$. So, fields with different
statistical behaviour are present in the same algebra, however the Hilbert
space remains separable.

We want to emphasize that relation of the type $\,\psi(x+\delta_x) =
U_{x+\delta_x}\,$ may be misleading, the latter element exists in the
crossed product only by Eq.(A.6), so that for the derivative one finds
\beqa 
\frac{\partial\psi(x)}{\partial x} := \lim_{\delta_x\ra 0}
\frac{\psi(x+\delta_x) - \psi(x)}{\delta_x} = \lim_{\delta_x\ra
0}\frac{1}{\delta_x}\left(V_{x_\delta} U_x - U_x\right)  = \no \\
\lim_{\delta_x\ra 0}\frac{1}{\delta_x}\left(e^{i\,2\pi\,\delta_x\,j(x)} -
1\right)U_x = 2\pi\,i\,j(x)U_x =: 2\pi\, i\,j(x) \psi(x).  
\eeqa 
This, together with (2.5) gives for the operators 
\beq 
i\psi_{f'} = \psi_{f}\, j_{\Theta'}.  
\eeq 
Note that in the crossed product, which can actually be considered as a
left $\A$-module, equations of motion (A.8), (A.9) appear (due to this
reason)  without an antisymmetrization, which was the case with the
functional realization, Eq.(4.2), but otherwise the result is the same.
Therefore the scaling sensitivity of the crossed product field algebra is
another manifestation of the quantum ``selection rule" for the value of
$\,\lambda\,$ in the Urgleichung (1.2c).

\end{document}